# Influence of oxygen pressure and aging on $LaAlO_3$ films grown by pulsed laser deposition on $SrTiO_3$ substrates

Jihwey Park[1], Yeong-Ah Soh[1,2, a)], Gabriel Aeppli[1], Adrian David[3], Weinan Lin[3], and Tom Wu[4]
[1]London Centre for Nanotechnology, University College London, London WC1H 0AH, United Kingdom
[2]Department of Materials, Imperial College London, London SW7 2AZ, United Kingdom
[3]Division of Physics and Applied Physics, Nanyang Technological University, Singapore 637371
[4]Physical Sciences and Engineering Division, King Abdullah University of Science and Technology, Thuwal 23955-6900, Saudi Arabia

The crystal structures of $LaAlO_3$ films grown by pulsed laser deposition on $SrTiO_3$ substrates at oxygen pressure of $10^{-3}$ mbar or $10^{-5}$ mbar, where kinetics of ablated species hardly depend on oxygen background pressure, are compared. Our results show that the interface between $LaAlO_3$ and $SrTiO_3$ is sharper when the oxygen pressure is lower. Over time, the formation of various crystalline phases is observed while the crystalline thickness of the $LaAlO_3$ layer remains unchanged. X-ray scattering as well as atomic force microscopy measurements indicate three-dimensional growth of such phases, which appear to be fed from an amorphous capping layer present in as-grown samples.
a) email: yeongahsoh@gmail.com

The interface between $LaAlO_3$ (LAO) and $SrTiO_3$ (STO) has attracted enormous attention since the discovery of its metallicity[1] even though both compounds are band insulators. Furthermore, recent findings of magnetism,[2] superconductivity,[3] and their coexistence[4] make this system even more popular. An intrinsic electronic reconstruction model in which electrons rearrange to avoid the so-called polar catastrophe[5] provides attractive explanations for novel properties such as a metal-insulator transition depending on the LAO thickness[6,7] or an applied electric field,[7] but the intrinsic model does not account for some observations.[8] The dependence of the properties on growth parameters in particular indicates that extrinsic factors play crucial roles.[1,2,9-17] An important growth parameter in pulsed laser deposition (PLD) is the background oxygen partial pressure, which has been believed to influence the conductivity by controlling the density of oxygen vacancies.[9-12] Recent studies, however, reveal that aspects of the interface structure such as intermixing and lattice distortion are also strongly influenced by the oxygen partial pressure during the LAO deposition.[15-17] So far, it has been found that an oxygen pressure high enough to reduce significantly the kinetic energy of growth species results in sharp interfaces,[16] but the relation between the interface roughening and the oxygen pressure in the region where the oxygen pressure hardly affects the kinetic energy of species still remains unexplored. In addition, cation intermixing across the interface has been suggested to be an important factor for the interface conductivity,[5,17-20] and therefore knowledge of the structural roughening depending on the oxygen pressure and sample age may help us to resolve the inconsistent observations of interface properties.

Surface x-ray scattering (SXS) is a powerful tool widely used to study the structure of thin films. SXS has been applied to investigate the interface structure between LAO and STO and has revealed lattice distortion and cation intermixing occurring at the interface depending on growth conditions.[18-22] Here we applied SXS to study the influence of the oxygen pressure on the structure of LAO/STO interfaces, and SXS data measured on LAO films grown at oxygen partial pressures of $10^{-3}$ and $10^{-5}$ mbar are compared. In this range of oxygen pressures, the kinetics of species ablated by lasers are almost independent of the oxygen pressure[15,16] , so similar roughnesses for the films were expected. However, we find that the films grown at the lower oxygen pressure have sharper boundaries than those at the higher pressure, independent of the film thickness. In addition, we observe a serious degradation of the film over time, consisting of significant roughening of the film morphology and formation of new phases, which may result in seemingly inconsistent experimental results, e.g. electrical transport, for the same sample as it ages.

STO and LAO have perovskite structures with (pseudo)cubic lattice constants of 3.905 Å and 3.79 Å, respectively. A set of LAO films with different nominal thicknesses were grown on $TiO_2$-terminated STO (001) substrates by PLD at oxygen partial pressure of $10^{-3}$ mbar and another set at $10^{-5}$ mbar. Two-dimensional layer-by-layer growth was confirmed by in-situ monitoring of reflection high-energy electron diffraction (RHEED) oscillations. The laser wavelength, fluence, and pulse repetition rate were 248 nm, 1.5 J/cm$^2$, and 1 Hz, respectively. The target-to-sample distance was 5 cm, the size of the ablated area was 6 mm$^2$, and the substrate temperature was held at 800 °C.. X-ray scattering measurements were performed at X20A and X18A beamlines of the National Synchrotron Light Source at Brookhaven National Laboratory in April and September 2011, respectively. Double Si (111) and Ge (111) crystals were used to obtain monochromatic x-rays at 8 keV at X18A and X20A, respectively. All measurements were carried out in air at room temperature using a standard four-circle diffractometer equipped with a scintillation counter. A Si (111) analyzer in front of the detector was used only at X20A.

Fig. 1 presents the reflectivity of 11 unit cell (uc) thick films measured in April 2011 and fitted assuming that the stoichiometry of the LAO layers is conserved, which is justified by carefully adjusting the laser fluence in the PLD growth to maintain the stoichiometric transfer of ablated species[23]. We can set a bound of 2% on the La deficiency of our samples based on the calibration of our PLD system done using X-ray photoelectron spectroscopy and the transport measurements carried out on our as-grown samples (data not presented here) that look very similar to the stoichiometric sample in Ref 23. The fits show that the LAO films have an additional layer at the top in addition to the layer with the expected thickness and electron density. The first layer directly on top of the substrate has the same electron density as that of bulk LAO[24] for both samples within experimental error, and its thickness is 39.3±0.3 Å for the $10^{-3}$ mbar-grown sample and 40.9±0.2 Å for the $10^{-5}$ mbar-grown sample, which correspond to the nominal thickness of 11 uc. The second layer has an electron density of 32±2 % of that expected in bulk LAO for both samples. The thickness of the second layer is 21.5±0.1 Å for the $10^{-3}$ mbar sample and 19.4±0.1 for the $10^{-5}$ mbar sample. The roughnesses of the surface and interfaces obtained by fitting the reflectivity data are summarized in Table I. The interface between the LAO layer and the STO substrate (interface II) is rougher than that between the additional layer and the LAO layer (interface I) for both samples. The roughnesses of the surface and the interface II for the $10^{-3}$ mbar sample are 1.8±1.1 Å larger than those for the $10^{-5}$ mbar sample while the roughnesses of the interface I are the same for both samples within the experimental error as seen in Table I.

Fig. 2 shows x-ray diffraction (XRD) intensities measured in April 2011 along (00L) and (10L) crystal truncation rods (CTRs), where momenta are denoted by the crystallographic axes of the STO substrate and its *c* axis is defined to be perpendicular to the interface. Laue oscillations originating from the finite thickness of the LAO layer are seen in addition to the STO and LAO Bragg peaks. The LAO layer thicknesses corresponding to these oscillations are 27±4 Å and 42.5±2 Å for the nominal thicknesses of 7 uc and 11 uc, respectively, which are largely independent of the oxygen pressure. The LAO layer thickness obtained from the Laue oscillations for the 11 uc films is the same as the thickness of the first layer directly on top of STO obtained from surface reflectivity. Since the Laue oscillations beside the Bragg peaks originate from interference of all crystal lattices within a finite thickness so that only crystalline regions contribute, whereas any abrupt change of electron density across the interface contributes to the reflectivity, we can conclude that the first layer directly on top of STO consists of crystalline LAO and the additional layer seen in the reflectivity data is amorphous. All films are fully strained, i.e., their in-plane lattice constants are the same as those of STO, which was confirmed by reciprocal space maps measured around the (103) reflection (not shown).

The Laue oscillations in Fig. 2 show strong asymmetric shapes. Larger intensities at the lower angle side of each Bragg peak than those with the same satellite orders at the larger angle side indicate that the LAO lattices along the surface normal direction are elongated near the interface, consistent with previous reports.[18,20-22] Here we did not characterize the lattice distortions quantitatively because of the lack of experimental data for reliable fitting. Nevertheless, the similar intensity ratios of the ±*n*th order satellite peaks near the Bragg positions indicate that the lattice distortion is largely insensitive to the oxygen pressure.

The oscillations far from the Bragg peak, by contrast, show clear distinction between the samples grown at different oxygen pressures. The oscillation for the sample grown at the oxygen pressure of $10^{-5}$ mbar is clearer than that of $10^{-3}$ mbar and this feature is independent of the LAO thickness. Blue lines in Fig. 3(b) show how the intensity profile of scattered x-rays evolves as a function of surface or interface roughness. We modeled the local atomic distributions of the surface and the LAO/STO interface by a Gaussian function in which its standard deviation corresponds to the roughness, as depicted in Fig. 3(a). In this model, therefore, the La and Al ions (or Sr and Ti) move together. In addition, lattice elongation of LAO along the surface normal direction is modeled by a power law plus a constant ($(x+a)^{-b}+c$), in which the out-of-plane lattice constant far from the interface is 3.73 Å,[21] and no deficiency of cations is assumed. The calculation shows that the intensity of high order satellites and the visibility of the oscillations decrease as the roughness increases. In addition, the oscillations of the satellites to the left of the Bragg peaks are stronger if the interface roughness is larger than the surface roughness, and vice versa. By comparing our experimental results to the calculations we can, therefore, infer that the boundaries of the $10^{-3}$ mbar-grown films are rougher than those of the $10^{-5}$ mbar-grown films and the interfaces are rougher than the surfaces for all the samples. Although the model is too simplified to provide quantitative agreement with the experimental data, the trends in roughness are consistent with the interface roughnesses obtained by fitting the x-ray reflectivity data of the 11 uc-thick LAO samples. This conclusion is consistent with the fact that Sr diffusion in STO is minimized at low oxygen pressures,[25] and the preservation of ordered $TiO_2$-terminated STO surface during the PLD growth improves the interface sharpness in low pressure grown samples.

Since interdiffusion of La and Sr is known to occur at LAO/STO interfaces, we considered the contribution of interdiffusion to the interface roughness, and incorporated population of Sr sites by La ions and vice versa into the same model as above. Based on the report[18] that the interdiffusion depth of La is larger by around one unit cell than that of Al, and that $La_{1-x}Sr_xTiO_3$ can, therefore, be formed around the interface, we assume a $TiO_2$ layer is above the interdiffusion-derived interface layer consisting of 50% LaO and 50% SrO that results in the formation of $La_{0.5}Sr_{0.5}TiO_3$. We model the depth profile of the ratio of LaO or SrO over the total rare earth oxide (LaO+SrO) by an error function with the roughness defined by its standard deviation. The XRD patterns obtained after incorporating composition roughness, ranging from 0 to 6 Å, introduced by interdiffusion in addition to the morphological roughness are qualitatively the same as before incorporating composition roughness, as shown in Fig. 3(b). Our calculated XRD patterns show that the influence of compositional roughness introduced by the interdiffusion of La and Sr to the XRD pattern is much smaller than that due to the roughness from morphology. Therefore, the contribution of the interdiffusion to the roughness measured with XRD is negligible and therefore, the XRD data mostly reflects the roughness due to morphology.

Fig 4 shows the XRD patterns measured in September 2011, five months after the first experiments. Strikingly, the results show dozens of additional peaks with different intensity profiles, indicating that the samples have undergone aging. The samples were stored at ambient atmosphere and temperature between the two experiments. The additional peaks are detected as long as the specular condition is satisfied, which indicates that the additional phases are crystalline with well-defined orientations. The full width at half maximum (FWHM) of the rocking curve of every additional peak was smaller than 0.1° for the LAO sample with 31 uc thickness grown at $10^{-5}$ mbar oxygen pressure. We have performed reciprocal line scans along the direction normal to the surface on nine LAO films with various thicknesses grown under the same conditions except for the oxygen partial pressure being either $10^{-3}$ or $10^{-5}$ mbar. 61 different positions of additional peaks have been observed and an additional peak at L = 2.931 seen in the first experiment became absent. Table II lists the samples that exhibit additional peaks and their corresponding L values and candidate materials composed of La, Sr, Ti, Al, or O that present Bragg reflections corresponding to the measured L values. Since there are many candidate materials we cannot specify exactly which materials have formed after the growth, beyond noting that there are many possibilities which researchers will need to be aware of. Despite the huge changes of the diffraction pattern after aging, the crystalline thickness obtained from the Laue oscillation remains essentially the same.

Atomic force microscopy (AFM) images show considerable changes of the surface morphology over time. An AFM image obtained on the 7 uc-thick LAO film grown under $10^{-5}$ mbar-oxygen-pressure in April 2011, just after the sample growth, shows terraces with unit cell step height, as presented in Fig. 5(a). This sharp surface degrades over time, as shown in Fig. 5(b) taken in April 2012. The height variation of the aged surface where the towers have heights of 5-50 nm is one order of magnitude larger than the original film thickness, 2.7 nm, even while we know from the x-ray data that the LAO layer on top of the STO remains flat and of the same thickness. All of our observations are therefore consistent with each other, and lead to the conclusion that the majority of the crystalline layer of the film remains unchanged with age, and it is the initially amorphous layer which provides the material from which the crystalline phases responsible for the tower-like surface structures in AFM images and the three-dimensional Bragg peaks grow.

Degradation of the conductivity at the LAO/STO interface over time has been reported and it has been proposed that oxygen diffusion induces the degradation by filling oxygen vacancies.[26] However, in that previous study, the structure of the films was not tracked, and based on our data we suspect that, besides the possible dynamics of oxygen-related defects, the increase of the resistivity can be attributed to the replacement of the amorphous, two-dimensional layer by much more textured crystalline phases. The observation that other phases are more easily formed from amorphous LAO films than for crystalline films[27] can explain well the fact that the resistivity of amorphous films were found to increase much more rapidly over time than that of crystalline films.[26]

In summary, we have compared LAO films grown at the oxygen pressures of $10^{-3}$ mbar or $10^{-5}$ mbar and found that the films at lower oxygen pressure have sharper boundaries. We observed strong evidence indicating that the films undergo aging, which may cause the degradation of the interface conductivity. The amorphous layer on top of the LAO layer appears to be at the root of the aging phenomena in LAO/STO heterostructures. Understanding such dynamic processes and the surface reactions involved will facilitate the development of strategies for the preservation of the desired physical properties of interface-based materials and devices.

We thank Christos Panagopoulos for discussions and EPSRC for funding the project.

REFERENCES

[1]A. Ohtomo and H. Y. Hwang, Nature **427**, 423 (2004).

[2]A. Brinkman, M. Huijben, M. VanZalk, J. Huijben, U. Zeitler, J. C. Maan,W. G. Van DerWiel, G. Rijnders, D. H. A. Blank, and H. Hilgenkamp, Nature Mater. **6**, 493 (2007)

[3]N. Reyren, S. Thiel, A. D. Caviglia, L. Fitting Kourkoutis, G. Hammerl, C. Richter, C. W. Schneider, T. Kopp, A.-S. Rüetschi, D. Jaccard, M. Gabay, D. A. Muller, J.-M. Triscone, and J. Mannhart, Science **317**, 1196 (2007).

[4]L. Li, C. Richter, J. Mannhart and R. C. Ashoori, Nature Phys. **7**, 762 (2011); J. A. Bert, B. Kalisky, C. Bell, M. Kim, Y. Hikita, H. Y. Hwang, and K. A. Moler, Nature Phys. **7**, 767 (2011).

[5]N. Nakagawa, H. Y. Hwang, and D. A. Muller, Nature Mater. **5**, 204 (2006).

[6]S. Thiel, G. Hammerl, A. Schmehl, C. W. Schneider, J. Mannhart, Sicence**313**, 1942 (2006)

[7]M.Huijben, G.Rijnders, D. H. A. Blank, S.Bals, S. VanAert, J.Verbeeck, G. VanTendeloo, A. Brinkman, and H.Hilgenkamp, Nature Mater. **5**, 556 (2006).

[8]H. Chen, A. M. Kolpak, and S. Ismail-Beigi, Adv. Mater.**22**, 2881 (2010).

[9]W. Siemons, G. Koster, H. Yamamoto, W. A. Harrison, G. Lucovsky, T. H. Geballe, D. H. A. Blank, and M. R. Beasley, Phys. Rev. Lett. **98**, 196802 (2007).

[10]G. Herranz, M. Basletić, M. Bibes, C. Carrétéro, E. Tafra, E. Jacquet, K. Bouzehouane, C. Deranlot, A. Hamzić, J. M. Broto, A. Barthélémy, and A. Fert, Phys. Rev. Lett. **98**, 216803 (2007).

[11]A. Kalabukhov, R. Gunnarsson, J. Börjesson, E. Olsson, T. Claeson, and D. Winkler, Phys. Rev. B **75**, 121404(R) (2007).

[12]M. Basletic, J.-L. Mauric, C. Carrétéro, G. Herranz, O. Copie, M.Bibes, É. Jacquet, K. Bouzehouane, S. Fusil, and A. Barthélémy, Nature Mater. **7**, 621 (2008).

[13]J.-L. Maurice, G. Herranz, C. Colliex, I. Devos, C. Carrétéro, A. Barthélémy, K. Bouzehouane, S. Fusil, D. Imhoff, É. Jacquet, F. Jomard, D. Ballutaud, M. Basletic, Europhys. Lett. **82**, 17003 (2008).

[14]G. Drera, G. Salvinelli, A. Brinkman, M. Huijben, G. Koster, H. Hilgenkamp, G. Rijnders, D. Visentin, and L. Sangaletti, arXiv:1211.5519 (2012).

[15]C. Aruta, S. Amoruso, R. Bruzzese, X. Wang, D. Maccariello, F. M. Granozio, and U. S. Di Uccio, Appl. Phys. Lett. **97**, 252105 (2010).

[16]W. S. Choi, C. M. Rouleau, S. S. A. Seo, Z. Luo, H. Zhou, T.T. Fister, J. A. Eastman, P. H. Fuoss, D. D. Fong, J. Z. Tischler, G. Eres, M. F. Chisholm, and H. N. Lee, Adv. Mater. **24**, 6423 (2012).

[17]A. Kalabukhov, A. B. Yu, I. T. Serenkov, V. I. Sakharov, J. Börjesson, N. Ljustina, E. Olsson, D. Winkler, and T. Claeson, Europhys. Lett. **93**, 37001 (2011).

[18]P. R. Willmott, S. A. Pauli, R. Herger, C. M. Schlepütz, D. Martoccia, B. D. Patterson, B. Delley, R. Clarke, D. Kumah, C. Cionca, and Y. Yacoby, Phys. Rev. Lett. **99**, 155502 (2007).

[19]S. A. Chambers, M. H. Englehard, V. Shutthanandan, Z. Zhu, T. C. Droubay, T. Feng, H. D. Lee, T. Gustafsson, E. Garfunkel, A. Shah, J. M. Zuo, and Q. M. Ramasse, Surf.Sci. Rep. **65**, 317 (2010).

[20]V. Vonk, J. Huijben, D. Kukuruznyak, A. Stierle, H. Hilgenkamp, A. Brinkman, and S. Harkema, Phys. Rev. B **85**, 045401 (2012).

[21]S. A. Pauli, S. J. Leake, B. Delley, M. Björck, C. W. Schneider, C. M. Schlepütz, D. Martoccia, S. Paetel, J. Mannhart, and P. R. Willmott, Phys. Rev. Lett. **106**, 036101 (2011).

[22]R. Yamamoto, C. Bell, Y. Hikita, H. Y. Hwang, H. Nakamura, T. Kimura, and Y. Wakabayashi, Phys. Rev. Lett. **107**, 036104 (2011).

[23]E. Breckenfeld, N. Bronn, J. Karthik, A. R. Damodaran, S. Lee, N. Mason, and L. W Martin, Phys. Rev. Lett. **110**, 196804 (2013).

[24]C. T. Chantler, J. Phys. Chem. Ref. Data **24**, 71 (1995); C. T. Chantler, J. Phys. Chem. Ref. Data **29**, 597 (2000).

[25]R. Moos and K. H. Hardtl, J. Am. Ceram. Soc. **80**, 2549 (1997).

[26]F. Trier, D.V. Christensen, Y.Z. Chen, A. Smith, M.I. Andersen, and N. Pryds, Solid State Ion. **230**, 12 (2013).

[27]M. He, G.-Z. Liu, W.-F. Xiang, H.-B. Lu, K.-J. Jin, Y.-L. Zhou, and G.-Z. Yang, Chin. Phys. Lett. **24**, 2671 (2007).

TABLE I. Roughnesses obtained from the reflectivity fits of the 11-uc thick LAO samples as shown in Fig. 1 (units in Å).

| position[a] | $10^{-3}$ mbar | $10^{-5}$ mbar |
|---|---|---|
| surface | 5.4±0.2 | 3.6±0.2 |
| interface I | 2.5±0.1 | 2.4±0.1 |
| interface II | 7.6±0.5 | 5.8±0.6 |

[a]Interface I and II represent the boundaries of the crystalline LAO layer with the additional amorphous top layer and the STO substrate, respectively.

TABLE II. Materials candidates for the additional x-ray diffraction peaks in Fig. 4.

| Number of Sample[a] | L | Candidates |
|---|---|---|
| April[b] | 2.931 | $LaAl_4$, $Ti_4O_7$, $Al_2Ti_7O_{15}$, $SrAl_4O_7$, $Sr_2Al_6O_{11}$, $La_2SrAl_2O_7$ |
| 1 | 0.760[c], 1.083, 1.284, 1.347, 1.390, 1.406, 1.526, 1.644, 1.654, 2.315, 2.576, 2.595, 2.597, 2.609 | $Al_2O_3$, $LaAl_4$, $SrAl_2$, $Sr_5Al_9$, $TiO$, $TiO_2$, $Ti_3O_5$, $Ti_7O_{13}$, $AlTiO_5$, $Al_2Ti_7O_{15}$, $La_2Ti_2O_7$, $La_4Sr_3O_9$, $La_4Ti_9O_{24}$, $SrAl_4O_7$, $SrTi_{11}O_{20}$, $Sr_2Al_6O_{11}$, $Sr_2Ti_6O_{13}$, $Sr_4Al_{14}O_{25}$, $Sr_4Ti_3O_{10}$, $Sr_4Ti_3O_{10}$, $Sr_{10}Al_6O_{19}$, $La_3Al_{15}Ti_5O_{37}$ |
| 2 | 0.752, 0.772, 0.806[c], 1.299, 1.519, 1.612, 1.630, 1.656, 1.834, 2.164, 2.444, 2.525 | $Al_2O_3$, $LaAl_4$, $SrAl$, $Sr_5Al_9$, $Sr_8Al_7$, $TiO_2$, $Ti_3O$, $Ti_6O$, $Ti_7O_{13}$, $Ti_8O_{15}$, $Ti_9O_{17}$, $LaAlO_3$, $LaTiO_3$, $La_2Ti_2O_7$, $La_4Ti_3O_{12}$, $La_4Ti_9O_{24}$, $La_5Ti_5O_{17}$, $SrAl_2O_4$, $SrTi_{11}O_{20}$, $Sr_2Al_6O_{11}$, $Sr_2Ti_6O_{13}$, $Sr_4Al_{14}O_{25}$, $Sr_{10}Al_6O_{19}$, $La_2SrAlO_7$, $La_3Al_{15}Ti_5O_{37}$ |
| 3 | 0.815, 1.623 | $La_9Ti_7O_{27}$, $SrAl_2O_4$, $SrAl_4O_7$, $Sr_4Al_{14}O_{25}$ |
| 4 | 0.861, 1.251, 1.290, 1.607, 1.804, 2.580 | $LaAl_4$, $La_2O_3$, $La_3Al_{11}$, $SrO$, $TiO_2$, $Ti_3O_5$, $Ti_5O_9$, $Ti_6O_{11}$, $Ti_7O_{13}$, $Ti_8O_{15}$, $La_4Sr_3O_9$, $Sr_2Al_6O_{11}$, $Sr_4Al_{14}O_{25}$, $SrAl_2O_4$, $SrTi_{11}O_{20}$, $Sr_4Al_{14}O_{25}$, $La_3Al_{15}Ti_5O_{37}$ |
| 5 | 1.220, 1,543, 1.585, 1.720, 2.439, 2.885, 2.961 | $LaAl$, $LaAl_4$, $La_3Al_{11}$, $SrAl_2$, $TiO_2$, $Ti_2O_3$, $Ti_4O_5$, $Ti_4O_7$, $Ti_6O$, $Ti_6O_{11}$, $Ti_8O_{15}$, $LaTiO_3$, $La_4Sr_3O_9$, $La_2Ti_2O_7$, $La_5Ti_5O_{17}$, $SrAl_2O_4$, $SrAl_4O_7$, $SrAl_{12}O_{19}$, $SrAl_{14}O_{25}$, $SrTi_{11}O_{20}$, $Sr_4Al_{14}O_{25}$, $Sr_{10}Al_6O_{19}$, $La_3Al_{15}Ti_5O_{37}$ |
| 6 | 2.500 | $LaA_{l4}$, $La_3Al_{11}$, $SrAl_{12}O_{19}$, $Sr_{10}Al_6O_{19}$ |
| 7 | 2.431 | $Al_2O_3$, $La_3Al_{11}$, $Ti_7O_{13}$, $Al_2TiO_5$, $La_4Sr_3O_9$, $La_4Ti_3O_{12}$, $La_5Ti_4O_{15}$, $La_9Ti_7O_{27}$, $SrTi_{11}O_{20}$, $La_3Al_{15}Ti_5O_{37}$ |
| 8 | 0.722[c], 1.530, 1.647, 1.945, 2.333 | $LaAl_4$, $La_3Al_{11}$, $TiO_2$, $Ti_2O$, $Ti_4O_7$, $Ti_5O_9$, $Ti_6O$, $Ti_6O_{11}$, $Ti_7O_{13}$, $La_2Ti_2O_7$, $SrAl_2O_4$, $SrTi_{11}O_{20}$, $Sr_2Al_6O_{11}$, $Sr_2Ti_6O_{13}$, $Sr_3Ti_2O_7$, $Sr_4Al_{14}O_{25}$, $Sr_4Ti_3O_{10}$, $Sr_7Al_{12}O_{25}$, $Sr_{10}Al_6O_{19}$, $La_3Al_{15}Ti_5O_{37}$ |
| 9 | 0.781, 0.974, 1.334, 1.443, 1.459, 1.501, 1.560, 1.667, 2.340, 2.667, 2.917, 3.120, 3.332 | $Al_2O_3$, $AlTi$, $AlTi_3$, $LaAl_4$, $La_3Al_{11}$, $SrAl$, $Sr_5Al_9$, $Sr_8Al_7$, $SrO_2$, $TiO$, $TiO_2$, $Ti_2O_3$, $Ti_3O_5$, $Ti_4O_5$, $Ti_4O_7$, $Ti_5O_9$, $Ti_6O_{11}$, $Ti_7O_{13}$, $Ti_8O_{15}$, $Ti_9O_{17}$, $Al_2TiO_5$, $Al_2Ti_7O_{15}$, $LaAlO_3$, $LaTiO_3$, $La_2TiO_5$, $La_2Ti_2O_7$, $La_4Sr_3O_9$, $La_4Ti_9O_{24}$, $La_5Ti_5O_{17}$, $SrAl_2O_4$, $SrAl_{12}O_{19}$, $Sr_2Al_6O_{11}$, $Sr_2TiO_4$, $Sr_2Ti_6O_{13}$, $Sr_3Al_2O_6$, $Sr_4Al_{14}O_{25}$, $Sr_4Ti_3O_{10}$, $Sr_{10}Al_6O_{19}$, $La_3Al_{15}Ti_5O_{37}$ |

[a]Number of samples represents how many samples show a peak at the (00L).
[b]All four samples measured in April 2011 show the additional peak.
[c]There is no candidate material corresponding to this L value which may come from materials having non-stoichiometric compositions.

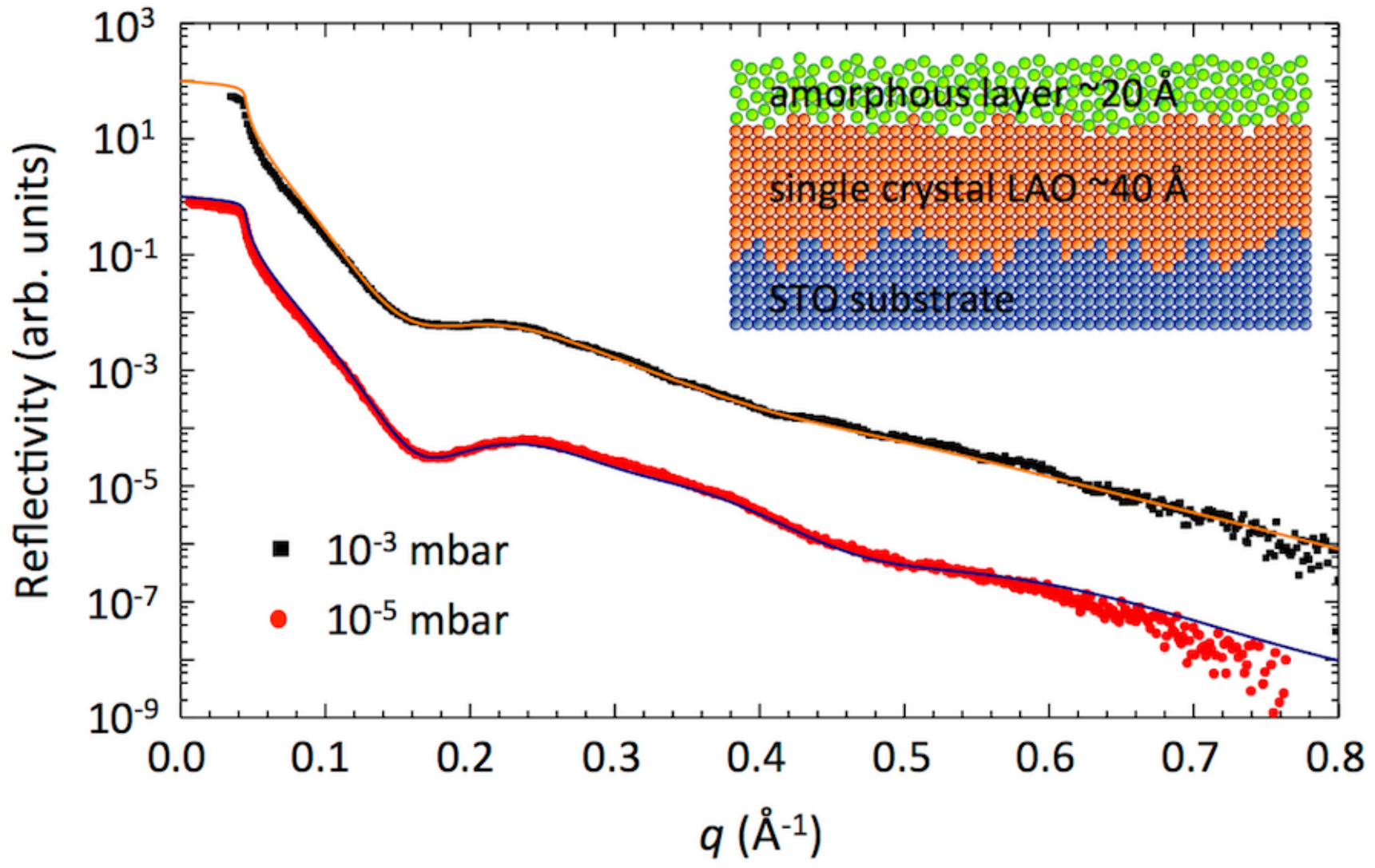

FIG. 1. X-ray reflectivity of the 11 uc-thick LAO films measured in April 2011. Black squares and red circles represent the samples grown at oxygen pressures of $10^{-3}$ and $10^{-5}$ mbar, respectively, and solid lines represent their corresponding fits. The curves have been shifted along the y-axis for clarity. The fits show the samples have two layers; the first layer from the STO substrate has a thickness consistent with 11 uc LAO and the same electron density as the bulk LAO within experimental errors while the second one has an electron density of 32±2 % of the bulk LAO for both samples and a thickness of 21.5±0.1 Å for the $10^{-3}$mbar sample and 19.4±0.1 for the $10^{-5}$mbar sample. The surface and the interface between the substrate and the first layer of the $10^{-3}$ mbar sample are rougher than those of the $10^{-5}$ mbar sample, and the roughness of the interface between the two layers is similar for both samples. Inset: Schematic structure of the fresh as-grown films.

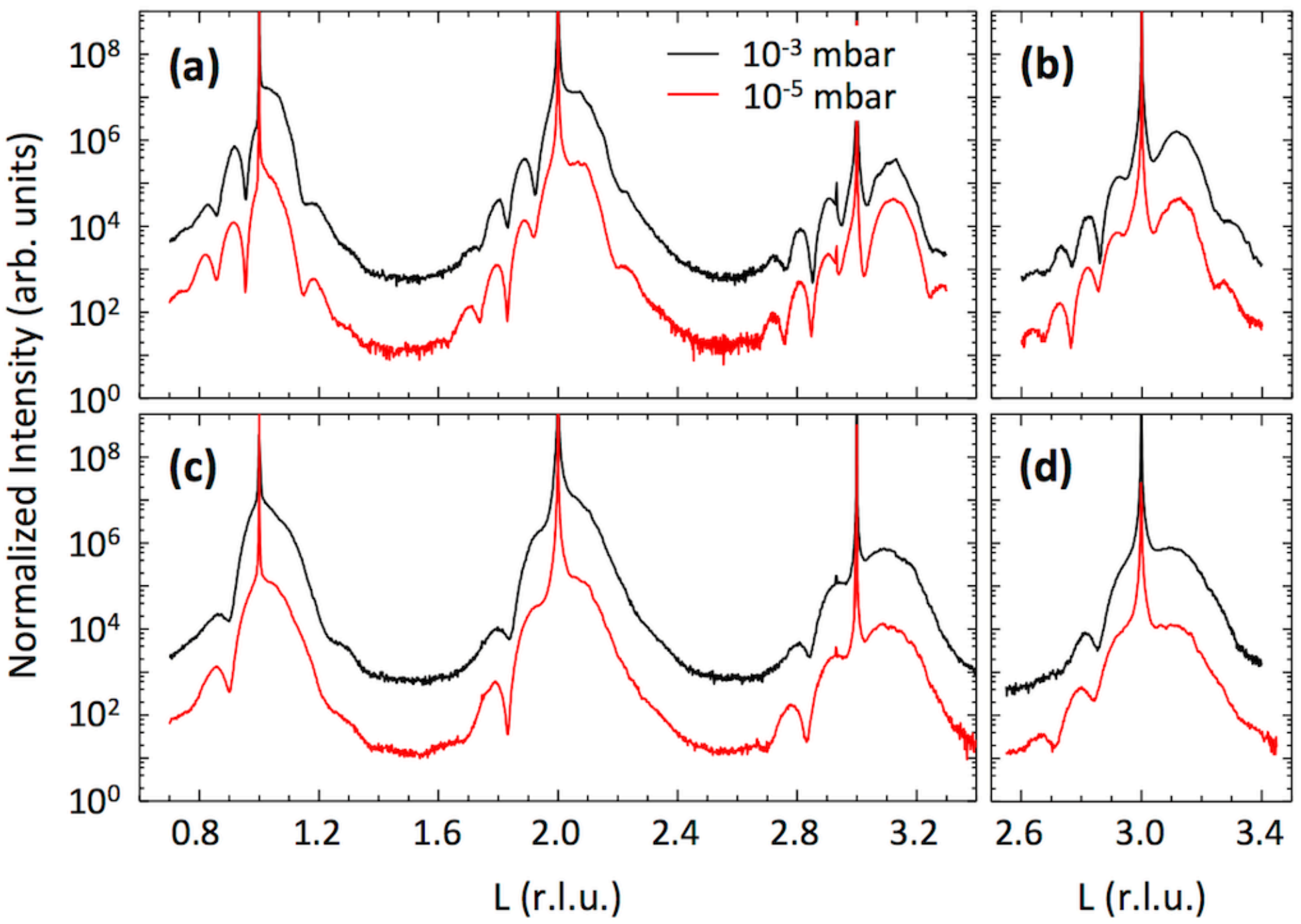

FIG. 2 XRD measurements in April 2011. Scattered intensities measured along the (00L) (a) and (10L) (b) directions of the films with a nominal LAO thickness of 11 uc. Panels (c) and (d) represent those for the films with nominal thickness of 7 uc. Black and red lines represent data on the samples grown at oxygen partial pressure of $10^{-3}$ mbar and $10^{-5}$ mbar, respectively. The curves have been shifted along the y-axis for clarity. Clearer oscillations for $10^{-5}$ mbar indicate that the films grown at $10^{-5}$ mbar oxygen pressure have sharper boundaries than those at $10^{-3}$ mbar.

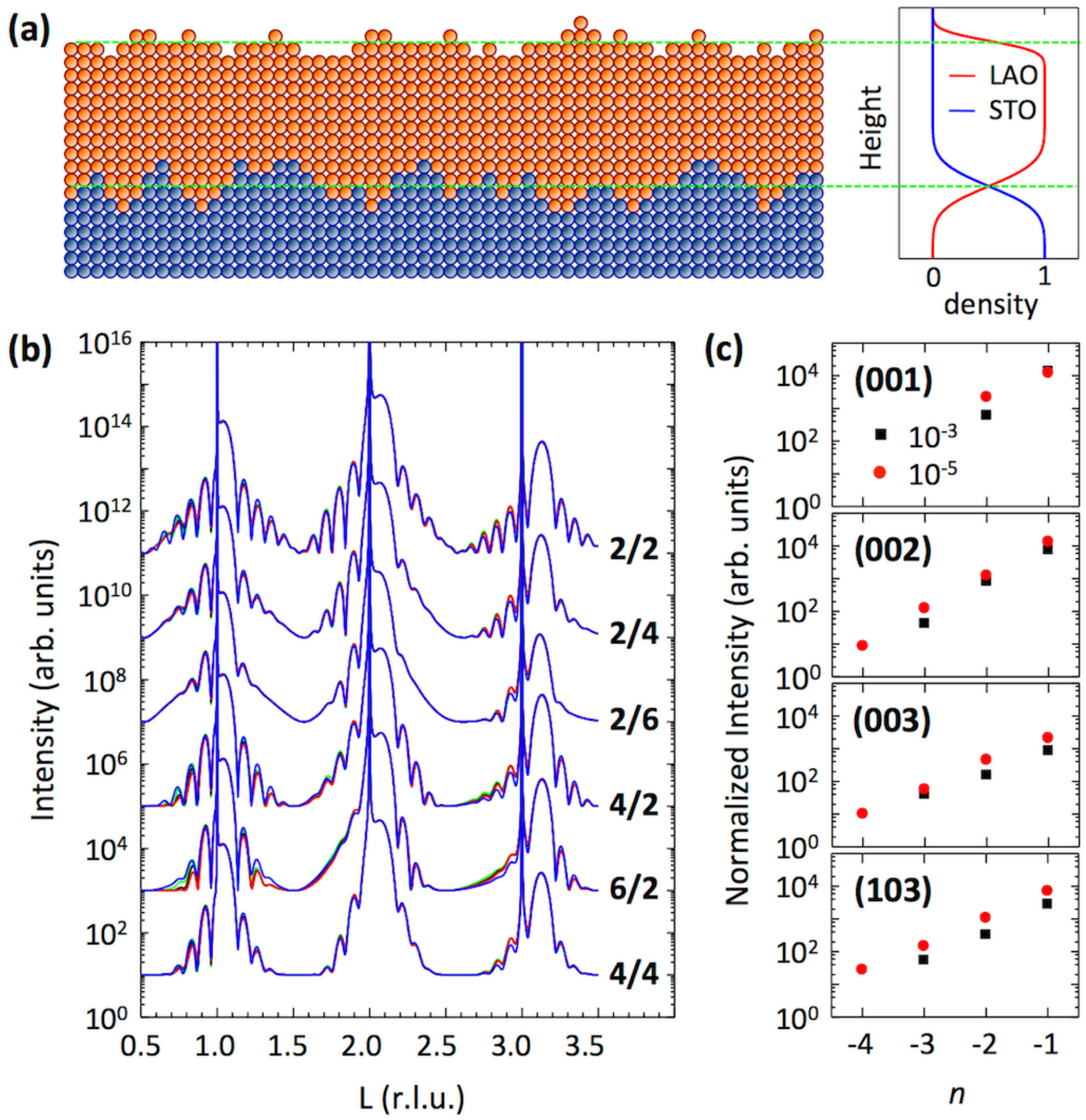

FIG. 3 (a) [left] Schematic structure for the calculation. Orange and blue circles represent LAO and STO unit cells, respectively. [right] Atomic density of LAO and STO as a function of the height along the direction normal to the surface. (b) Calculated XRD patterns with the kinematical theory with an average LAO thickness of 11 uc. In the calculation we assumed (1) a Gaussian distribution of the surface and interface heights, (2) elongation of the out-of-plane lattice constant of LAO near the interface according to a power law, and (3) absence of any deficiency of cations and oxygen. Blue lines represent the limit where there is not interdiffusion, and green, black, and red lines are for interdiffusion widths of 2, 4, and 6 Å (standard deviations in Gaussian error function), respectively. The numbers on the right side of the curves represent the standard deviations of the Gaussian functions, i.e., roughnesses, in units of Å; the first number is for the surface and the second for the interface. (c) Measured peak height of the *n*th order satellite of the 11 uc thick LAO film. A constant background has been subtracted. Black squares and red circles represent the $10^{-3}$ and $10^{-5}$ mbar oxygen pressures, respectively. In every case, the peak height of the $10^{-5}$ mbar film is higher than that of the $10^{-3}$ mbar one.

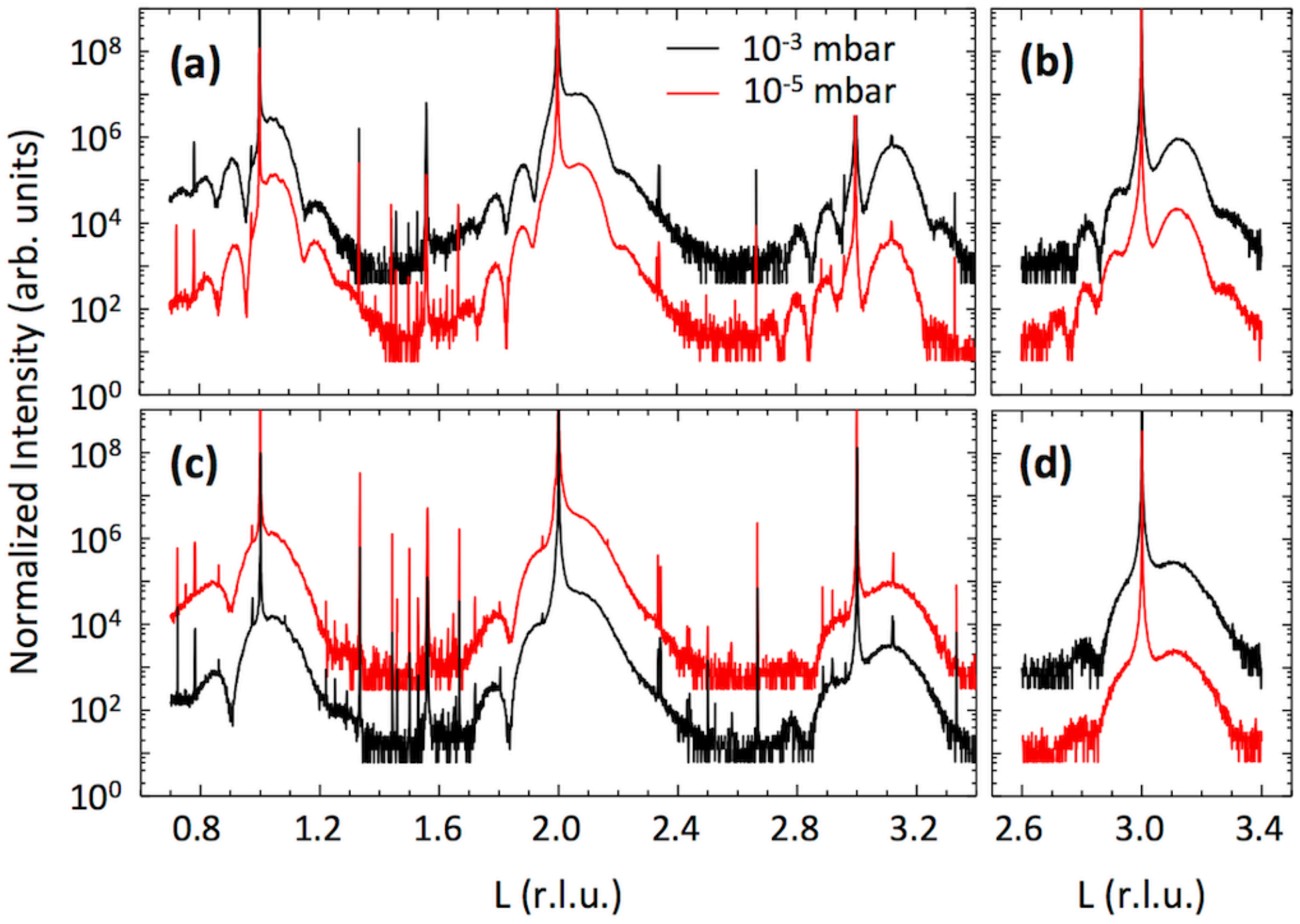

FIG. 4 XRD measurements in September 2011 of the same samples as in Fig 2. Scattered intensities measured along the (00L) (a, c) and (10L) (b, d) directions on the films with a nominal LAO thickness of 11 uc and 7 uc, respectively. Black and red lines represent the oxygen pressures of $10^{-3}$ mbar and $10^{-5}$ mbar, respectively. The curves have been shifted along the y-axis for clarity. Dozens of additional peaks satisfying the specular condition are observed, which indicates the presence of additional crystalline phases with well-defined orientations.

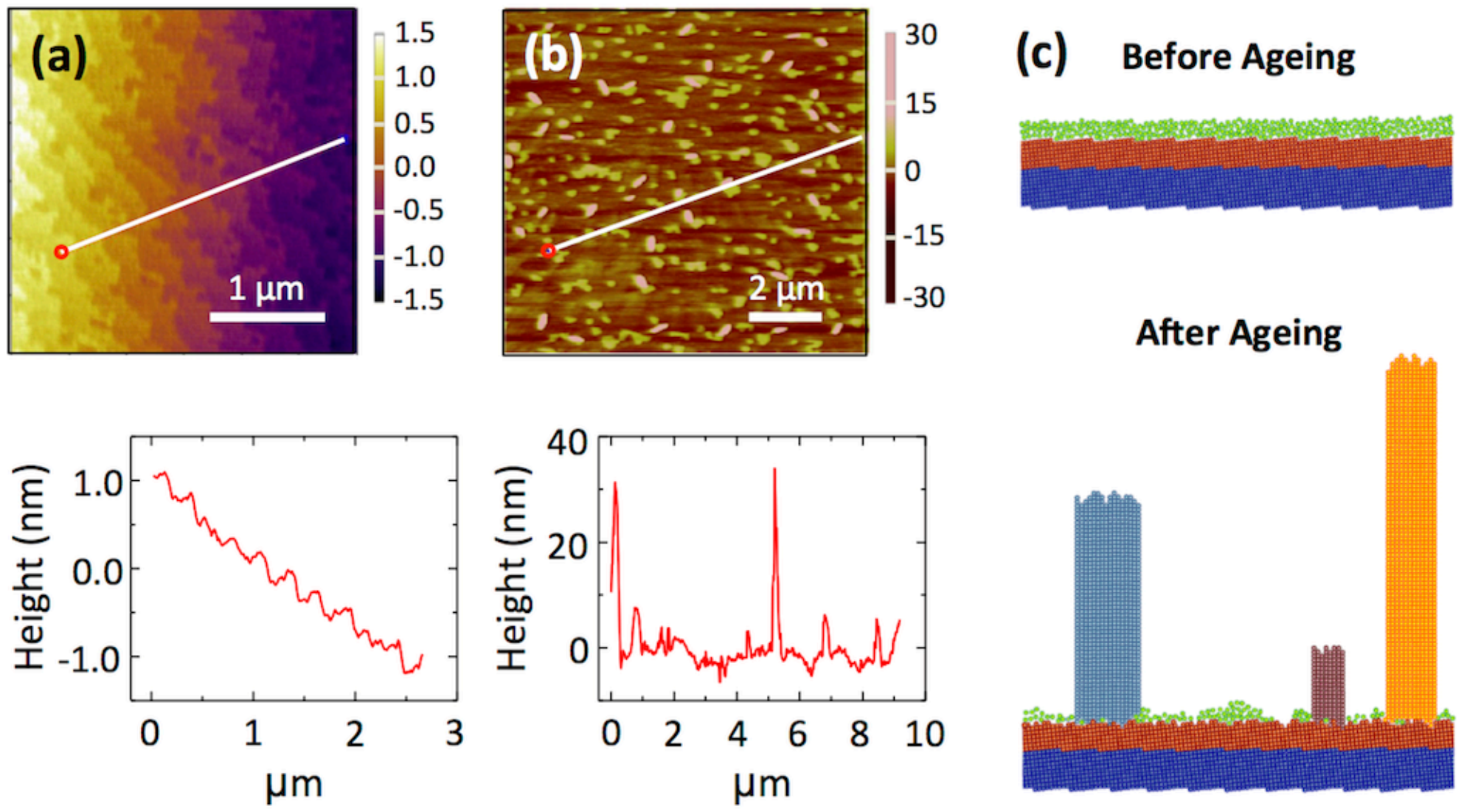

FIG 5. AFM images (higher panels) and line profiles (lower panels) of the 7 uc thick LAO sample grown at $10^{-5}$ mbar oxygen pressure measured in April 2011 (a) and in April 2012 (b). While the image taken just after the growth shows terraces with unit cell steps, the image taken one year later shows that the film surface has undergone ageing and became inhomogeneous. (c) Schematic diagram of the LAO/STO heterostructure before and after ageing. The scale parallel to the film surface is one order larger than that normal to the surface.